\documentclass[twocolumn,prl,showpacs,amsmath,amssymb]{revtex4-1}

\usepackage{graphicx}% Include figure files
\usepackage{dcolumn}% Align table columns on decimal point
\usepackage{bm,color}% bold math   \textcolor{red} ON

\begin{document}

\preprint{APS/}

\title{Large-scale shell-model analysis of the neutrinoless $\beta\beta$ decay of $^{48}$Ca}

\author{Y. Iwata$^{1}$}
 \email{iwata@cns.s.u-tokyo.ac.jp}
\author{N. Shimizu$^{1}$}
\author{T. Otsuka$^{1,2,3,4}$} 
\author{Y. Utsuno$^{1,5}$}
\author{J. Men\'endez$^{2}$}
\author{M. Honma$^{6}$}
\author{T. Abe$^{2}$}

 \affiliation{
$^{1}$Center for Nuclear Study, The University of Tokyo, 113-0033 Tokyo, Japan \\
$^{2}$Department of Physics, The University of Tokyo, 113-0033 Tokyo, Japan \\
$^{3}$National Superconducting Cyclotron Laboratory, Michigan State University, East Lansing, MI 48824, USA\\
$^{4}$Instituut voor Kern- en Stralingsfysica, Katholieke Universiteit Leuven, B-3001 Leuven, Belgium\\
$^{5}$Advanced Science Research Center, Japan Atomic Energy Agency, Tokai, 319-1195 Ibaraki, Japan  \\
$^{6}$Center for Mathematical Sciences, University of Aizu, 965-8580 Fukushima, Japan
}

\date{\today}% It is always \today, today,

\begin{abstract}
We present the nuclear matrix element for the neutrinoless double-beta decay of $^{48}$Ca
based on large-scale shell-model calculations including two harmonic oscillator shells
($sd$ and $pf$ shells).
The excitation spectra of $^{48}$Ca and $^{48}$Ti, and the two-neutrino double-beta decay of $^{48}$Ca are reproduced in good agreement to experiment.
We find that the neutrinoless double-beta decay nuclear matrix element
is enhanced by about 30\% compared to $pf$-shell calculations.
This reduces the decay lifetime by almost a factor of two.
The matrix-element increase is mostly due to pairing correlations associated with cross-shell $sd$-$pf$ excitations.
We also investigate possible implications for heavier neutrinoless double-beta decay candidates.
\end{abstract}

\pacs{23.40.-s, 21.60.Cs, 23.40.Hc, 27.40.+z}% PACS, the Physics and Astronomy
                             % Classification Scheme.and
\maketitle

The observation of neutrino oscillations established the massive nature of neutrinos
almost two decades ago~\cite{98fukuda}. 
Despite great progress in neutrino physics in recent years~\cite{14gonzalez},
some fundamental properties are still unknown,
like the Dirac or Majorana neutrino nature (whether they are their own antiparticle),
or the absolute neutrino mass-scale and hierarchy.
The first question would be answered with the detection of neutrinoless double-beta 
($0\nu\beta\beta$) decay.
In this lepton-number violating process, a nucleus decays into its isobar
with two less neutrons and two more protons, emitting two electrons and no (anti)neutrinos.
Several international collaborations are running experiments to measure this process
~\cite{14exo,13kamlandzen,13gerda,15cuore}
or plan to do it in the near future~\cite{14candles,14majorana,14next,15supernemo,15cobra,15sno+},
and have set impressive lower-limits for the $0\nu\beta\beta$ decay lifetimes,
$T^{0\nu}_{1/2}>10^{25}$y, for the most favourable cases.

In addition, $0\nu\beta\beta$ decay can determine the absolute neutrino masses and hierarchy
if the nuclear matrix element (NME) of the transition, $M^{0\nu}$, is accurately known.
The lifetime of the decay reads~\cite{08avignone}
\begin{equation}  
\left[T_{1/2}^{0 \nu}\left(0^+_i\rightarrow 0^+_f\right)\right]^{-1}
= G^{0 \nu} |M^{0 \nu}|^2  \left( \frac{\langle m_{\beta \beta} \rangle}{m_e} \right)^2,  
\end{equation}
with $0^+_i$ ($0^+_f$) the initial (final) state, $G^{0\nu}$ a well-known phase-space factor~\cite{12kotila},
and ${\langle m_{\beta \beta}\rangle}$ a combination of the absolute neutrino masses
and the neutrino mixing matrix (the electron mass $m_e$ is introduced by convention).

Calculated NME values, however, differ by factors of two or three
depending on the theoretical nuclear structure approaches used.
This uncertainty severely limits the potential capability to determine the
absolute neutrino masses with $0\nu\beta\beta$ decay.
Among the NME calculations, shell-model results~\cite{09menendez,10horoi,13senkov}
are typically at the lower end,
and it has been argued that this may be due to the relatively small configuration space
that can be accessed by present shell-model codes~\cite{12vogel}.
On the other hand, within the configuration space where the calculation is performed,
the shell model can include various additional correlations
compared to other approaches that yield larger NME values~\cite{08caurier,14menendez,15menendez}, like the quasiparticle random-phase approximation (QRPA)
~\cite{13simkovic,15hyvarinen,15terasaki},
the interacting boson model (IBM)~\cite{15barea},
the energy density functional (EDF)~\cite{13vaquero,15yao},
or the generator coordinate method (GCM)~\cite{14hinohara}.

The doubly-magic $^{48}$Ca
is the lightest isotope considered in $\beta\beta$ decay searches,
including the CARVEL~\cite{05zdesenko}, CANDLES~\cite{14candles,08umehara,15kishimoto},
and NEMO-III~\cite{11barabash} experiments.
Its $\beta\beta$ decay into $^{48}$Ti is ideally suited for shell-model calculations,
which are very successful in this mass region for a wide variety of observables~\cite{05caurier}.
In fact, the two-neutrino double-beta ($2\nu\beta\beta$) decay lifetime
was predicted by a shell-model calculation~\cite{90caurier}
in very good agreement with the subsequent experimental detection~\cite{96balysh}. 

In this Letter we present an improved calculation of the $0\nu\beta\beta$ decay NME for $^{48}$Ca
based on the large-scale shell model in two harmonic oscillator shells
($sd$ and $pf$ shells).
This significantly expands previous shell-model studies performed in the $pf$ shell~\cite{08caurier,09menendez,10horoi,13senkov},
increasing the number of single-particle orbitals from four to seven.
We use the $M$-scheme shell-model code KSHELL~\cite{13shimizu},
and allow up to 2$\hbar \omega$ $sd$-$pf$ cross-shell excitations.
The dimension of the largest calculation ($^{48}$Ti) is $2.0\times 10^{9}$.

\begin{figure}[t]
 \includegraphics[width=1.\columnwidth]{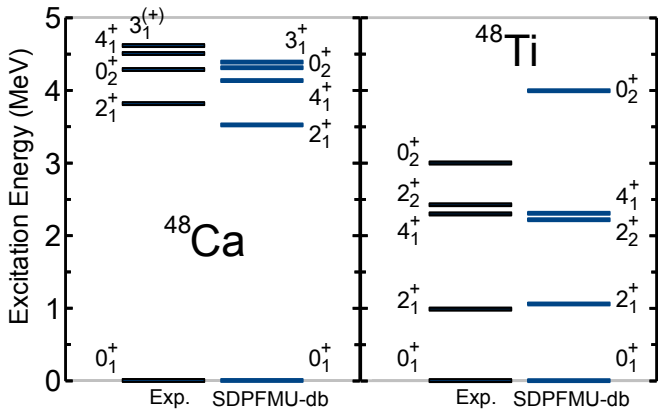}
 \caption{(color online)
Excitation spectra of $^{48}$Ca and $^{48}$Ti.
The lowest five positive-parity states~\cite{nudat}
are compared to $sdpf$ calculations with the SDPFMU-db interaction.
} 
 \label{spectra}
 \end{figure}

We use the shell-model SDPFMU effective interaction~\cite{12utsuno},
which describes well the shell evolution and the spectroscopy
of neutron-rich nuclei in the upper $sd$ shell.
The $pf$-shell part of this interaction is based on the GXPF1B interaction,
which accounts very successfully for the spectroscopy of $pf$-shell nuclei~\cite{08honma,05honma}. 
While the SDPFMU interaction works reasonably well, a slightly revised one, SDPFMU-db, is introduced by reducing the shell gap of $^{40}$Ca to 5.8 MeV so as to reproduce the observed $0^+_2$ level of $^{48}$Ca. 
The two-proton transfer reaction experiment~\cite{86videbaek} shows a large cross section to the $0_2^+$ state of $^{48}$Ca, suggesting sizable proton excitations from the $sd$ shell.
The $0_2^+$ state obtained with the SDPFMU-db interaction shows 1.64 protons in the $pf$ shell consistently with this property, whereas the SDPFMU result finds only 0.22.
The new SDPFMU-db interaction thus gives an improved description compared to SDPFMU.

Figure~\ref{spectra} shows the excitation spectra of $^{48}$Ca and $^{48}$Ti
obtained with SDPFMU-db, which are in good agreement to experiment.
The SDPFMU spectra is generally of similar quality, with the $0^+_2$ level of $^{48}$Ca too high by 200 keV.
In contrast, a $pf$-shell calculation with GXPF1B gives the $0^+_2$ level in $^{48}$Ca $1.3$~MeV higher than the experimental one.
For the $0^+_2$ state in $^{48}$Ti, the $sdpf$-shell calculation with SDPFMU-db gives $1.0$~MeV higher excitation energy than experiment, probably due to missing $4\hbar\omega$ excitations.
The $2\hbar\omega$ components in the ground states of $^{48}$Ca and $^{48}$Ti
are 22\% and 33\% for SDPFMU-db (14$\%$ and 20$\%$ for SDPFMU).
Such sizable $2\hbar \omega$ excitations suggest that these interactions in the $sdpf$-configuration space capture sufficiently well cross-shell $sd$-$pf$ excitations.

\begin{figure}[t]
 \includegraphics[width=1.\columnwidth]{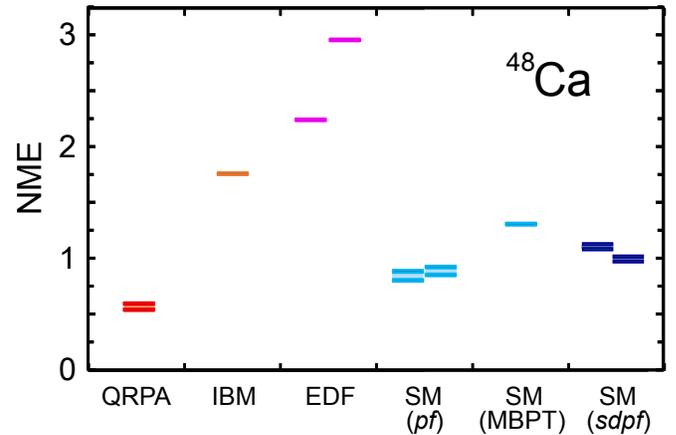}
 \caption{(color online)
Comparison of NME values for the $^{48}$Ca $0\nu\beta\beta$ decay.
The present shell-model results in the $sdpf$ space (SM $sdpf$: left~SDPFMU-db, right~SDPFMU) are compared to
$pf$-shell results (SM $pf$: left~\cite{13senkov}, right~\cite{09menendez}),
$pf$-shell result plus a perturbative calculation of the effect of orbitals
outside the $pf$ shell (SM MBPT)~\cite{14kwiatkowski},
QRPA~\cite{13simkovic}, IBM~\cite{15barea}, and EDF (left: non-relativistic~\cite{13vaquero}, right: relativistic~\cite{15yao}) calculations.
The range between double horizontal bars covers results including different type of short-range correlations
(Argonne, CD-Bonn, UCOM~\cite{05roth}) and without them.
} 
\label{nme_comp}
%%SM SDPF 0.96-1.18
%%SM PF 0.85-0.92 (menendez)
%%SM PF 0.80-0.88 (senkov)
%%SM MBPT 1.30
%%QRPA 0.54-0.59
%%IBM 1.75
%%NR EDF 2.23
%R EDF 2.94
\end{figure}

First we study the $2\nu\beta\beta$ decay of $^{48}$Ca.
We calculate the Gamow-Teller $\beta^+$ and $\beta^-$ strengths,
and compare them to experiments for the energy range up to 5 MeV~\cite{07grewe,09yako}, so that we can extract the appropriate quenching factor $q$ of the $\sigma\tau$ operator for each calculation.
We find $q=0.71$ for both $sdpf$ interactions,
and $q=0.74$ for the $pf$-shell interaction, in accordance with previous $pf$-shell studies~\cite{05caurier}.
The similar quenching factor shows that it does not depend on missing $sd$-$pf$ correlations.
Then we calculate $2\nu\beta\beta$ decay matrix elements
by summing contributions from 100 virtual $1^+$ intermediate states in $^{48}$Sc,
and obtain $M^{2\nu}=0.051$ $(0.045)$~MeV$^{-1}$ with the SDPFMU-db (SDPFMU) interaction,
in good agreement with experiment, $M^{2\nu}=0.046 \pm 0.004$~MeV$^{-1}$~\cite{15barabash}.
In the $pf$-shell calculation with GXPF1B the result is very similar, $M^{2\nu}=0.052$~MeV$^{-1}$,
reflecting low sensitivity to the size of the shell-model configuration space in $2\nu\beta\beta$ decay.
This is in contrast to the high sensitivity observed in Ref.~\cite{13horoi}.
The difference arises because in the present calculations all spin-orbit partners are always included.

We then calculate the $^{48}$Ca $0\nu\beta\beta$ decay NME in the $sdpf$ space including up to $2\hbar\omega$ configurations.
It is given in the closure approximation as~\cite{08avignone}
\begin{align}
M^{0 \nu}
 = \langle 0^+_f | \hat{O}^{0\nu} | 0^+_i \rangle
 = M^{0\nu}_{GT}-\frac{g_V^2}{g_A^2}M^{0\nu}_{F}+M^{0\nu}_{T},
\end{align}
with Gamow-Teller ($M^{0\nu}_{GT}$), Fermi ($M^{0\nu}_{F}$) and tensor ($M^{0\nu}_{T}$)
terms classified according to the spin structure of the operator.
The vector and axial coupling constants are taken to be $g_V=1$ and $g_A=1.27$, respectively.
We set the closure parameter to $\langle E \rangle=0.5$~MeV, found appropriate in the $pf$-shell calculation of Ref.~\cite{13senkov}.
We consider the inclusion of Argonne- and CD-Bonn-type short range correlations~\cite{09simkovic}.
Two-body current contributions to the transition operator~\cite{11menendez} are not included.
The possible quenching of the $\sigma\tau$ operator in $0\nu\beta\beta$ decay is the matter of discussion~\citep{12vogel}, because compared to $2\nu\beta\beta$ decay the momentum transfer is larger, and the virtual intermediate states of the transition include additional multipolarities. 
Therefore, similarly to most previous calculations, we do not quench the $\sigma\tau$ operator for $0\nu\beta\beta$ decay.
A detailed discussion on the $0\nu\beta\beta$ decay operator $\hat{O}^{0\nu}$ can be found in Ref.~\cite{10horoi}.

\begin{table*}[t]
\caption{NME value for the $^{48}$Ca $0\nu\beta\beta$ decay.
The $pf$-shell calculation with GXPF1B is compared to the $sdpf$ $2\hbar\omega$ results
obtained with the SDPFMU-db and SDPFMU interactions.
Total values ($M^{0 \nu}$) are shown together with Gamow-Teller ($M_{GT}^{0 \nu}$),
Fermi ($M_{F}^{0 \nu}$) and Tensor ($M_{T}^{0 \nu}$) parts.
Argonne- and CD-Bonn-type short-range correlations (SRC) are considered.
}
\label{nme_2hw}
\begin{center} 
\begin{tabular*}{\textwidth}{@{\extracolsep{\fill}}l|c c c c |c c c c|c c c c}
\hline \hline     
 & \multicolumn{4}{c}{GXPF1B} & \multicolumn{4}{c}{SDPFMU-db} & \multicolumn{4}{c}{SDPFMU}
 \\
 \hline    
 SRC  &  $M_{GT}^{0 \nu}$ & $M_{F}^{0 \nu}$ & $M_{T}^{0 \nu}$ & $M^{0 \nu}$    
      &  $M_{GT}^{0 \nu}$ & $M_{F}^{0 \nu}$ & $M_{T}^{0 \nu}$ & $M^{0 \nu}$   
      &  $M_{GT}^{0 \nu}$ & $M_{F}^{0 \nu}$ & $M_{T}^{0 \nu}$ & $M^{0 \nu}$   
       \\ \hline
 None    & 0.776 & $-0.216$ & $-0.077$ & 0.833 
         & 0.997 & $-0.304$ & $-0.067$ & 1.118
         & 0.894 & $-0.291$ & $-0.068$ & 1.007   \\
 CD-Bonn & 0.809 & $-0.233$ & $-0.074$ & 0.880
         & 1.045 & $-0.327$ & $-0.065$ & 1.183
         & 0.939 & $-0.313$ & $-0.065$ & 1.068   \\
 Argonne & 0.743 & $-0.213$ & $-0.075$ & 0.801
         & 0.953 & $-0.300$ & $-0.065$ & 1.073
         & 0.852 & $-0.288$ & $-0.068$ & 0.963  \\ 
 \hline  \hline
 \end{tabular*} 
\end{center}
\end{table*}		

The calculated values of NME are shown in Table~\ref{nme_2hw}.
The Gamow-Teller and Fermi parts,
$M_{GT}^{0 \nu}$ and $M_{F}^{0 \nu}$, are enhanced in the $2\hbar\omega$ calculations
by about $20\%-40\%$ compared to the $pf$-shell calculations.
The largest values are given by the SDPFMU-db interaction,
which allows a stronger mixing of $2\hbar\omega$ configurations in the mother and daughter nuclei.
The tensor contribution, $M_{T}^{0 \nu}$, is almost unaffected by enlarging the configuration space.
The 10\% difference between the NME values obtained with
the two $sdpf$ shell-model interactions is similar
to the uncertainty obtained with different $pf$-shell interactions~\cite{10horoi}.
The sensitivity to short-range correlations is about 10\%.
Using the closure parameter $\langle E \rangle=7.72$~MeV of Refs.~\cite{09menendez,10horoi},  the NME value is reduced by around 5\%.

Additional correlations beyond the $sd$-$pf$ space are potentially relevant for the $^{48}$Ca NME. To evaluate its effect, we have performed a $2\hbar\omega$ calculation including the $pf$ and $sdg$ shells, using the interaction from Ref.~\cite{12utsuno}, which describes well negative parity states in neutron-rich calcium isotopes (sensitive to $pf$-$sdg$ excitations). We find a small 5\% change in the NME compared to the $pf$-shell result, consistent with the small cross-shell $pf$-$sdg$ excitations (about 2\%) in $^{48}$Ca and $^{48}$Ti. This suggests that the $sd$-$pf$ space captures the most relevant correlations beyond the $pf$ shell for the $^{48}$Ca NME.

Figure~\ref{nme_comp} compares different NME calculations for $^{48}$Ca.
The total NME value in the $sdpf$ configuration space, $M^{0\nu}=0.96 - 1.18$,
is about 30\% larger than the $pf$-shell GXPF1B result or other shell-model $pf$-shell values $M^{0\nu}=0.78 - 0.92$~\cite{09menendez,10horoi,13senkov}.
This enhancement has important consequences for $^{48}$Ca $0\nu\beta\beta$ decay experiments,
as the decay lifetime is almost halved.
The present NME value is 15\% smaller than the result obtained by a $pf$-shell calculation including perturbatively the effect of the orbitals outside the $pf$ configuration space, $M^{0\nu}=1.30$ ~\cite{14kwiatkowski}.
In contrast, Fig.~\ref{nme_comp} shows that the present NME value is considerably smaller than IBM~\cite{15barea},
non-relativistic~\cite{13vaquero} or relativistic~\cite{15yao} EDF values,
and significantly larger than the QRPA result~\cite{13simkovic}.

\begin{figure} [t]
 \includegraphics[width=\columnwidth]{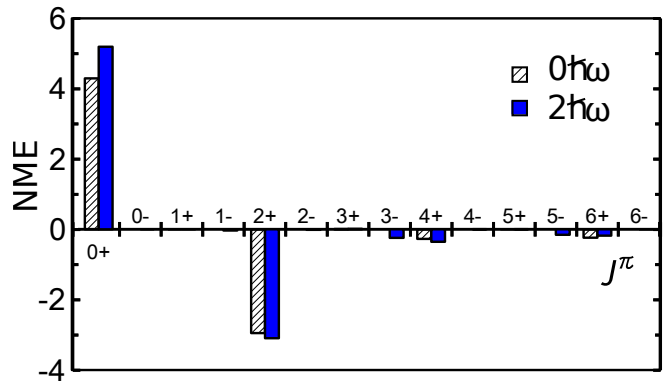}
 \caption{(color online)
NME decomposition in terms of
the angular momentum and parity, $J^{\pi}$, of the pair of decaying neutrons, 
Eq.~(\ref{decomp1}). 0$\hbar \omega$ (GXPF1B)
and 2$\hbar \omega$ (SDPFMU-db) results are compared, without short-range correlations.
} 
  \label{Jp decomp}
\end{figure}	

In the following we analyse the NME to understand the mechanisms responsible for the enhancement found in the $2\hbar\omega$ calculations,
and explore possible implications for heavier $0\nu\beta\beta$ decay candidates.
The operator for NME can be decomposed in terms of the angular momentum
and parity, $J^{\pi}$, to which the two-decaying neutrons are coupled~\citep{12vogel}:
\begin{equation}  \label{decomp1} 
M^{0 \nu} =  
\sum_J ~ \langle 0^+_f |  \sum_{i \le j,~k \le l}M_{ij,kl}^J [({\hat a}^{\dagger}_i {\hat a}^{\dagger}_j)^J 
({\hat a}_k {\hat a}_l)^J]^0 | 0^+_i \rangle, 
 \end{equation}
where $i,j,k,l$ label single-particle orbitals.
This decomposition is shown in Fig.~\ref{Jp decomp}
for $0\hbar \omega$ ($pf$) and $2 \hbar \omega$ ($sdpf$) calculations.
The leading contribution to $0\nu\beta\beta$ decay comes from $0^+$-coupled pairs,
while other $J^{\pi}$ combinations suppress the NME.
Figure~\ref{Jp decomp} shows that the main difference
between the $0\hbar \omega$ and $2 \hbar \omega$ results
is a 20\% increase in the contributions of $0^+$ pairs.
In addition only the $2 \hbar \omega$ calculation allows for negative-parity pairs,
but its contribution is small.
As also suggested in Ref.~\cite{14brown}, these findings indicate that the NME is enhanced by the pairing correlations, which induce $0^+$-pair excitations, introduced by the additional $sd$-shell orbitals.
  
\begin{figure}[t]  
 \includegraphics[width=\columnwidth]{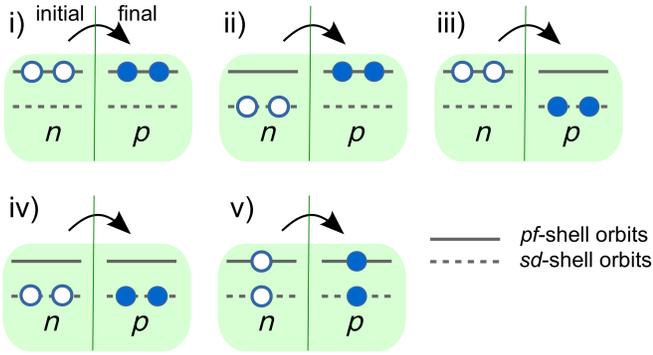}
\caption{(color online)
Diagrams associated with the NME decomposition in Eq.~(\ref{eq:diagrams}),
classified in terms of the $sd$- or $pf$-shell orbitals
occupied by the decaying neutrons (open circles) and created protons (filled circles).
Initial (final) stands for $^{48}$Ca ($^{48}$Ti).
Diagrams i-v correspond to
${\mathcal M}_{1}^{0 \nu}-{\mathcal M}_{5}^{0 \nu}$, respectively.
}
\label{diagrams} 
 \end{figure}

We further decompose the NME in terms of the orbitals ($sd$ or $pf$ shell)
occupied by the two $^{48}$Ca neutrons and two $^{48}$Ti protons involved in the decay:
\begin{equation} \label{eq:diagrams}
\begin{array}{ll}
M^{0 \nu}  = {\mathcal M}_{1}^{0 \nu}+{\mathcal M}_{2}^{0 \nu}+{\mathcal M}_{3}^{0 \nu} +{\mathcal M}_{4}^{0 \nu}+{\mathcal M}_{5}^{0 \nu} ,
\end{array}
\end{equation}
with the ${\mathcal M}^{0 \nu}$ components, sketched in Fig.~\ref{diagrams}, defined as
\begin{align}
{\mathcal M}_{1}^{0 \nu} =&
\langle 0_f^+  | \hat{O}^{0\nu}\left(p_{pf}~p_{pf};n_{pf}~n_{pf}\right)| 0_i^+ \rangle, \nonumber \\ 
{\mathcal M}_{2}^{0 \nu} =&
\langle 0_f^+  | \hat{O}^{0\nu}\left(p_{pf}~p_{pf};n_{sd}~n_{sd}\right)| 0_i^+ \rangle, \nonumber 
\\
{\mathcal M}_{3}^{0 \nu} =&
\langle 0_f^+  | \hat{O}^{0\nu}\left(p_{sd}~p_{sd};n_{pf}~n_{pf}\right)| 0_i^+ \rangle, \nonumber 
\\
{\mathcal M}_{4}^{0 \nu} =&
\langle 0_f^+  | \hat{O}^{0\nu}\left(p_{sd}~p_{sd};n_{sd}~n_{sd}\right)| 0_i^+ \rangle, \nonumber 
\\
{\mathcal M}_{5}^{0 \nu} =&
\langle 0_f^+  | \hat{O}^{0\nu}\left(p_{sd}~p_{pf};n_{sd}~n_{pf}\right)| 0_i^+ \rangle,
\end{align}
where $n_{i}$ ($p_{i}$) stands for neutrons (protons) in the $i$ shell
of $^{48}$Ca ($^{48}$Ti).
Table~\ref{table:diagrams} shows the different components
in Eq.~(\ref{eq:diagrams}) for the SDPFMU-db $2\hbar\omega$ calculation,
as well as their decomposition in terms of the $J^{\pi}$ of the decaying neutron pair
[cf.~Eq.~(\ref{decomp1})].
${\mathcal M}_{1}^{0 \nu}$, the only term allowed in the $0\hbar\omega$ calculation,
is very similar in the $pf$ and $sdpf$ spaces.
On the contrary, ${\mathcal M}_{2}^{0 \nu}$, ${\mathcal M}_{3}^{0 \nu}$
and ${\mathcal M}_{4}^{0 \nu}$ require $2\hbar\omega$ excitations
in the mother and/or daughter nuclei (see Fig.~\ref{diagrams}).
In fact, these terms are responsible for
the enhancement of the NME in the $sdpf$ configuration space.
Table~\ref{table:diagrams} shows that, for ${\mathcal M}_{2}^{0 \nu}$, ${\mathcal M}_{3}^{0 \nu}$ and ${\mathcal M}_{4}^{0 \nu}$, the contribution of $0^+$ pairs is dominant,
about three times larger in magnitude than the other $J^{\pi}$ pairs.
This is in contrast to ${\mathcal M}_{1}^{0 \nu}$, or $pf$-shell calculations, 
where the contribution of the 0$^+$ terms is 30\% larger than the other $J^{\pi}$ pairs.
These results confirm that the pairing correlations inducing neutron and proton cross-shell $sd$-$pf$ excitations are responsible for the enhancement of the NME.

\begin{table}[t]
\caption{NME decomposition of Eq.~(\ref{eq:diagrams}),
for a $sdpf$ $2\hbar\omega$ SDPFMU-db calculation without short-range correlations.
The total value is shown along with the contributions of $J^{\pi}=0^+$ and all remaining pairs.}
\begin{center}
  \begin{tabular*}{\columnwidth}{@{\extracolsep{\fill}}l r r r r r r r r} \hline \hline
& ${\mathcal M}_{1}^{0 \nu}$ & ${\mathcal M}_{2}^{0 \nu}$ & ${\mathcal M}_{3}^{0 \nu}$ & ${\mathcal M}_{4}^{0 \nu}$ &  ${\mathcal M}_{5}^{0 \nu}$    \\ \hline
 Total   &  $0.915$ &  $0.168$ & $0.269$ &  $0.220$ & $-0.454$  \\
 $J^{\pi}=0^+$   &  $4.193$ & $0.364$ &  $0.379$ &  $0.255$ &  $0.000$  \\
 $J^{\pi}=0^-,J>0$    & $-3.278$ & $-0.196$  & $-0.109$  & $-0.035$ & $-0.454$  \\ 
\hline  \hline
 \end{tabular*}
\end{center}
\label{table:diagrams}
\end{table}

The remaining term ${\mathcal M}_{5}^{0 \nu}$ requires the two nucleons being in different orbitals
(see Fig.~\ref{diagrams}, diagram v).
These two neutrons cannot be coupled to $J^{\pi} = 0^+$,
and are not involved in the $0^+$ pair contributions.
They instead produce strong cancellations, as shown in Table~\ref{table:diagrams}, consistently with the $J^{\pi} \ne 0^{+}$ contributions in Fig.~\ref{Jp decomp}.

The above discussion suggests that the enlargement of the model space produces two competing mechanisms to be considered in all $0\nu\beta\beta$ decays. 
On the one hand, additional pairing correlations in the mother and daughter nuclei, enhanced by two-particle--two-hole (2p-2h) excitations with respect to the original configuration space, increase the NME values, as seen in ${\mathcal M}_{1}^{0 \nu}-{\mathcal M}_{4}^{0 \nu}$ for the $^{48}$Ca decay.
On the other hand, excitations in the initial and final nuclei outside the original space can increase $J^{\pi} \ne0^+$ contributions as well.
Assuming that these follow the same trends as in Fig.~\ref{Jp decomp}, this second mechanism will reduce the NME value, as seen in ${\mathcal M}_{5}^{0 \nu}$ for $^{48}$Ca. 
Important contributions come from one-particle--one-hole (1p-1h) excitations. 
For the $^{48}$Ca decay, however, 1p-1h excitations always change parity and do not contribute to $0^+$ ground states, and this mechanism remains rather modest.

For heavier nuclei, these two competing effects need to be calculated in detail.
While pairing correlations are most important for $0\nu\beta\beta$ decay, 1p-1h type excitations have smaller unperturbed energy difference than 2p-2h excitations, and can be sizable. 
The balance between the two mechanisms will determine the NME. 
For example,  Ref.~\cite{13horoi} found a 35\% smaller NME value for $^{136}$Xe when including up to 1p-1h excitations into the missing spin-orbit partners in the original shell-model configuration space. 
In contrast, Ref.~\cite{08poves}  found a 20\% increase in the $^{82}$Se and $^{136}$Xe NME values when considering 2p-2h excitations.
A related competition between opposite-sign contributions was very recently suggested in Ref.~\cite{15brown} for $^{76}$Ge.

Finally, we estimate the NME beyond $2\hbar\omega$ $sd$-$pf$ excitations.
An exact diagonalization in the full $sdpf$ configuration space
is not feasible with present computing capabilities.
However this space can be handled in a seniority-zero approximation,
that is, in a basis with all nucleons coupled in like-particle $J^{\pi}=0^+$ pairs.
In a given configuration space the NME is maximum in this limit,
as higher seniority components only reduce their value~\cite{08caurier}.
A full $sdpf$ seniority-zero calculation with SDPFMU-db,
performed with the $J$-coupled code NATHAN~\cite{05caurier},
shows that components beyond $2\hbar\omega$ excitations are negligible
(less than 0.5\%) in both $^{48}$Ca and $^{48}$Ti.
That is, $N\hbar\omega$ excitations ($N>2$) only contribute to high-seniorities,
thus they can only reduce the NME.
This implies that the $sdpf$ pairing correlations enhancing $0\nu\beta\beta$ decay
are completely captured by the $2\hbar\omega$ configurations included in the present calculations,
and consequently the results obtained in this work provide an upper-bound for the NME value in the full $sdpf$ configuration space.

In summary, we have carried out large-scale shell-model calculations of $^{48}$Ca and $^{48}$Ti,
for the first time including up to $2\hbar\omega$ excitations in the $sdpf$ space.
The excitation spectra of $^{48}$Ca and $^{48}$Ti, and the $2\nu\beta\beta$ decay of $^{48}$Ca are reproduced in good agreement to experiment.
We find different sensitivities to the configuration-space size in $\beta\beta$ decays;
while the $2\nu\beta\beta$ decay NME is similar in the $pf$ and $sdpf$ shells,
the $0\nu\beta\beta$ decay NME increases by about 30\% to $M^{0\nu}\approx 1.1$.
The NME enhancement, which almost halves the associated decay life time, is due to cross-shell $sd$-$pf$ pairing correlations.
A seniority analysis shows that pairing effects in the $sdpf$ space are completely captured by the $2\hbar\omega$ calculations, so that the present result suggests an upper value for the NME in the full $sdpf$ space. 

Correlations outside the $sd$-$pf$ space have been evaluated to be small. Beyond present shell-model capabilities, they can be estimated with MBPT~\cite{14kwiatkowski} or GCM~\cite{15menendez,14hinohara} techniques, complementing the present result.
Further efforts are needed to set a more definitive value for the $^{48}$Ca $0\nu\beta\beta$ decay NME, for instance by further enlarging the model space, improving the closure approximation, introducing two-body currents and/or a renormalization of the operator for the model space.
Future plans include calculating NMEs
for heavier $0\nu\beta\beta$ decay candidates in extended shell-model configuration spaces. 
For these isotopes, competition between 1p-1h and pairing-like 2p-2h excitations in the present context will be of much interest, and their subtle balance should be evaluated precisely to obtain reliable NMEs. \\

This work was supported in part by Grants-in-Aid for Scientific Research (23244049, 25870168, $26\cdot04323$, 15H01029). 
It was supported also in part by HPCI Strategic Program (hp150224) and CNS-RIKEN joint project for large-scale nuclear structure calculations.
J. M. was supported by an International Research Fellowship from JSPS.
Numerical calculations were carried out at FX10 (The University of Tokyo), K computer (RIKEN AICS), and COMA (University of Tsukuba).

\end{document}